\documentclass{aastex62}

\shorttitle{C-1.4 Class Flare and Associate Jet}
\shortauthors{Solanki R. et al.}

\begin{document}
\title{C-1.4 Class Flare and An Associated Peculiar Coronal Jet}

\correspondingauthor{Ritika Solanki}
\email{ritikas.rs.phy15@iitbhu.ac.in}

\author{Ritika Solanki}
\affil{Department of Physics, Indian Institute of Technology (BHU), Varanasi - 221005, India.}

\author{Abhishek K. Srivastava}
\affil{Department of Physics, Indian Institute of Technology (BHU), Varanasi - 221005, India.}

\author{B. N. Dwivedi}
\affil{Department of Physics, Indian Institute of Technology (BHU), Varanasi - 221005, India.}

\begin{abstract}
Using HINODE/XRT, GOES, SDO/AIA observations, we study a compact C-1.4 class flare outside a major sunspot of AR 12178 on 4 October 2014. This flare is associated with a peculiar coronal jet, which is erupted in two stages in the overlying corona above the compact flaring region. At the time of flare maximum, the first stage of the jet eruption occurs above the flare energy release site, and thereafter in the second stage its magneto-plasma system interacts with the overlying distinct magnetic field domain in its vicinity to build further the typical jet plasma column.

\end{abstract}

\section{Introduction} \label{sec:intro}
Solar flares are magneticlly driven transients, which can release huge amount of energy in few minutes to few tens of minutes (Shibata \& Magara 2011). Generally flares occur in strong magnetic field regions (i.e., active regions), but sometimes they also occur in spotless regions (Dodson \& Hedeman 1970). Flare process may emit whole electromagnetic spectrum, energetic particles, and can accompanied with various physical processes, e.g., chromospheric evaporation, plasmoid generation, waves \& oscillations, jet/surge eruption, etc.  In the present paper, we briefly describe a compact C-class flare that gives rise a peculiar two stage jet eruption.
\begin{figure*}
\mbox{
\includegraphics[width=3in]{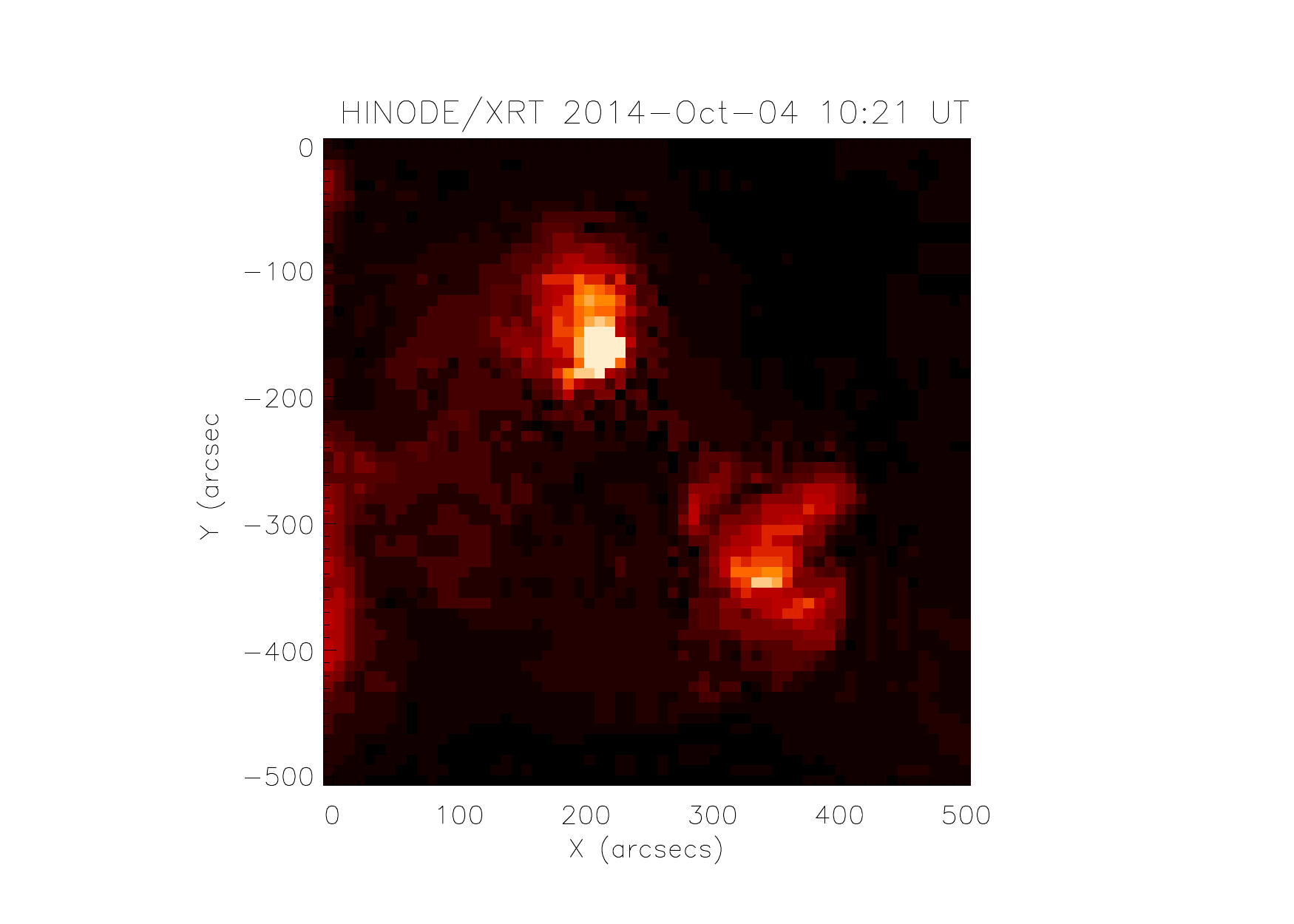}
\hspace{-1.5cm}
\includegraphics[width=3in]{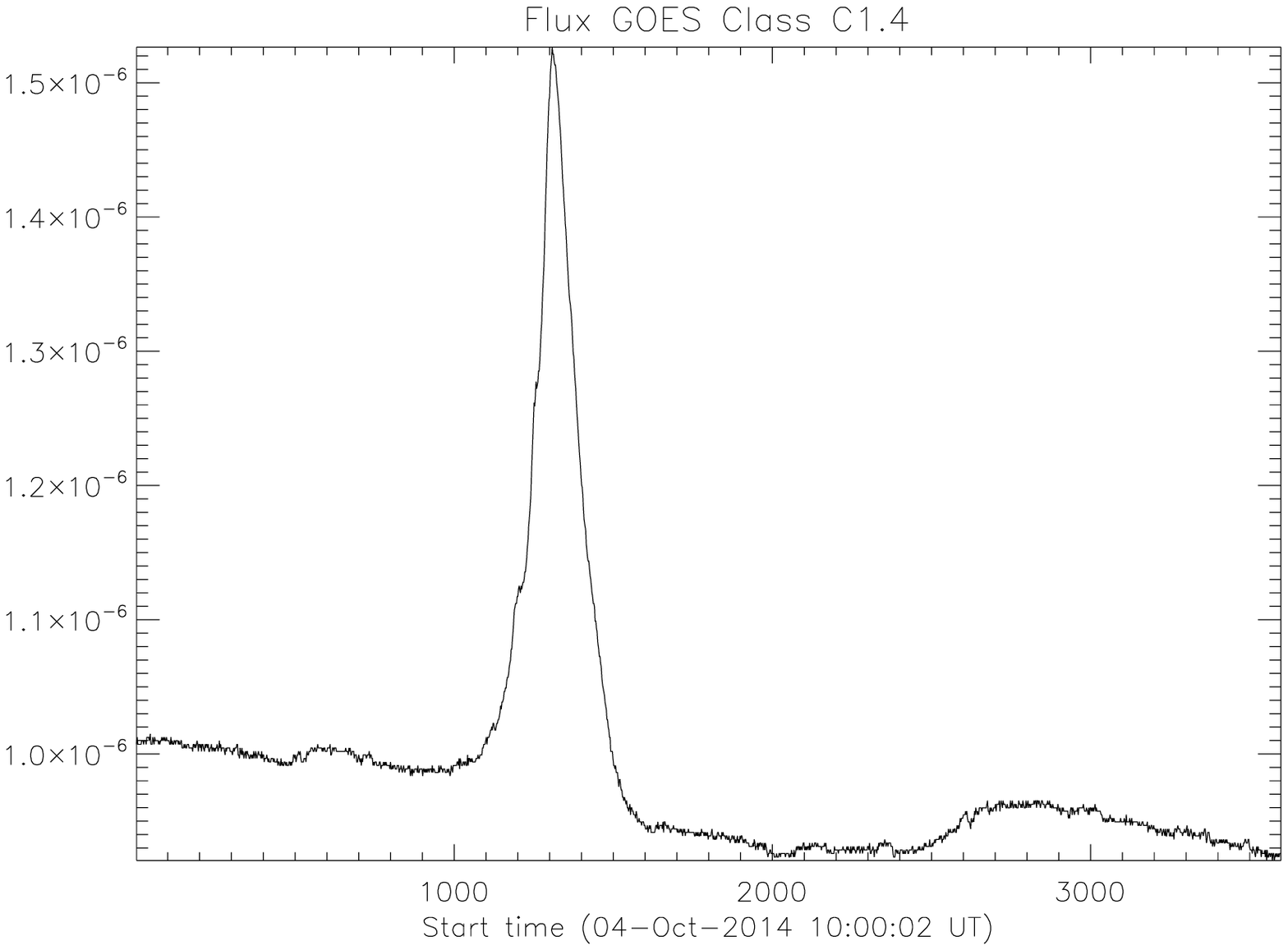}
}
\caption{Left panel: HINODE/XRT X-ray image shows a compact C1.4 class solar flare. Right panel: GOES light curve for this flare.}
\end{figure*}
\section{Observational Data and its Analyses}
For this paper, we use observational data from X-Ray Telescope (XRT) onboard Hinode (Golub et al.2007; Kosugi et al. 2007) and Atmospheric Imaging Assembly (AIA) (Lemen et al. 2012) onboard Solar Dynamics Observatory (SDO) (Pesnell, Thompson, \& Chamberlin 2012) in three EUV wavelengths 304\AA~, 171\AA~ and in 211\AA~. XRT observes solar plasma with temperatures between 1 and 10 MK with 1\arcsec pixel size. AIA has 0.6\arcsec pixel resolution and it operates with 12 sec cadence in EUV channels. We also use GOES data to show the evolution of the compact C1.4-class solar flare. 
\begin{figure*}
\vspace{-0.1cm}
\begin{center}
\mbox{
\includegraphics[width=6.0cm,height=5cm]{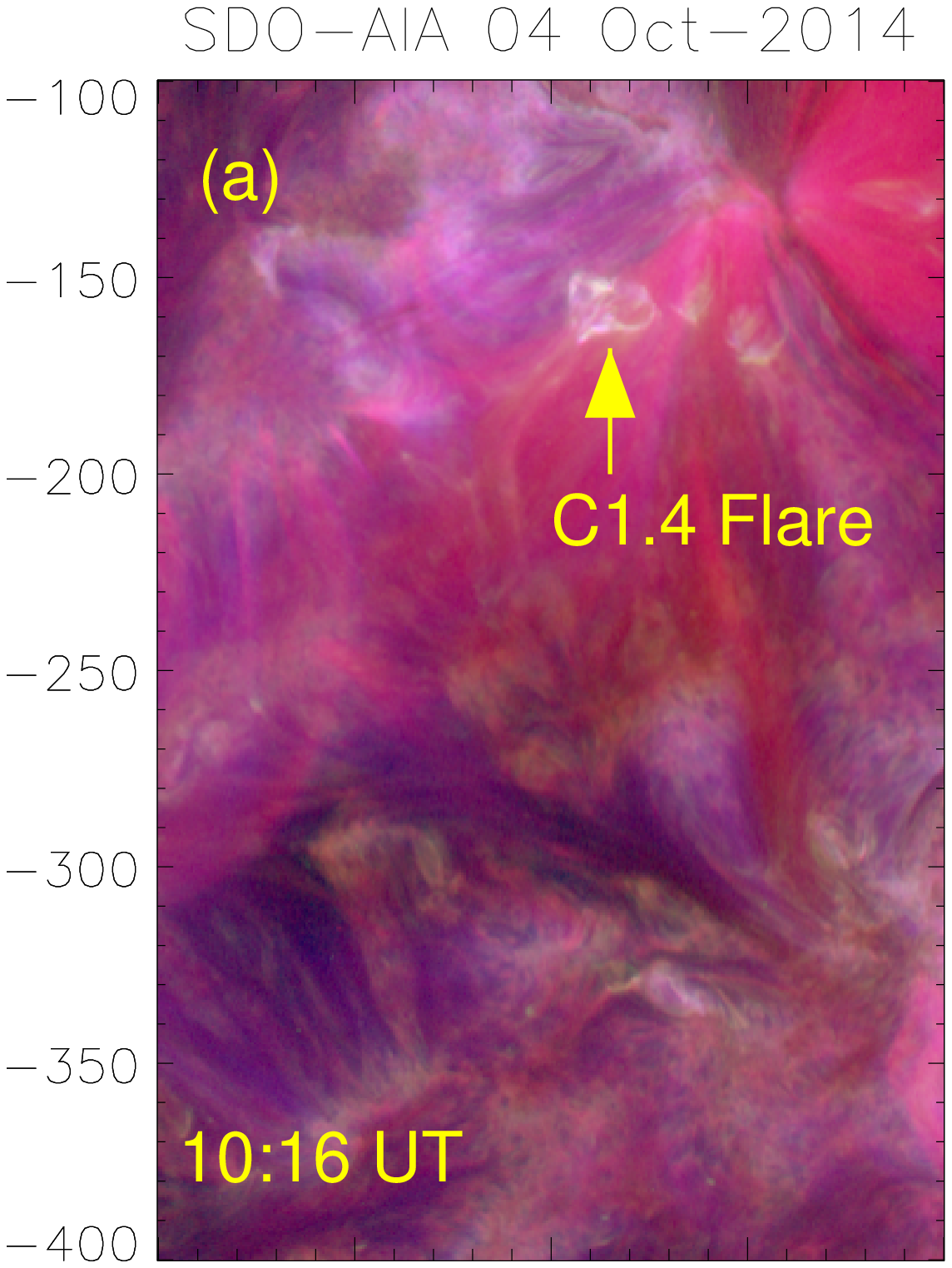}
\hspace{-3.0cm}
\includegraphics[width=6.0cm,height=5cm]{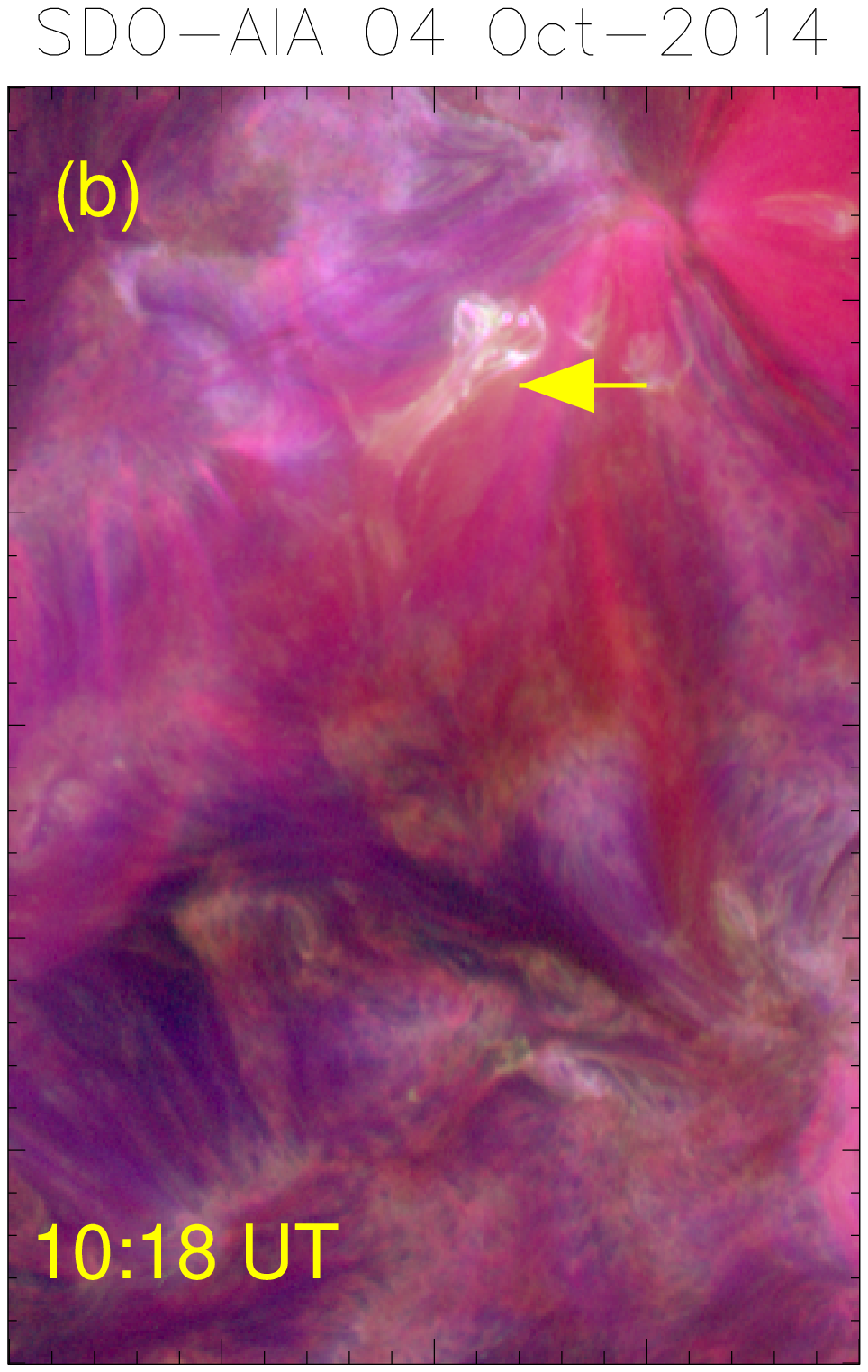}}
\mbox{
\includegraphics[width=6.0cm,height=5cm]{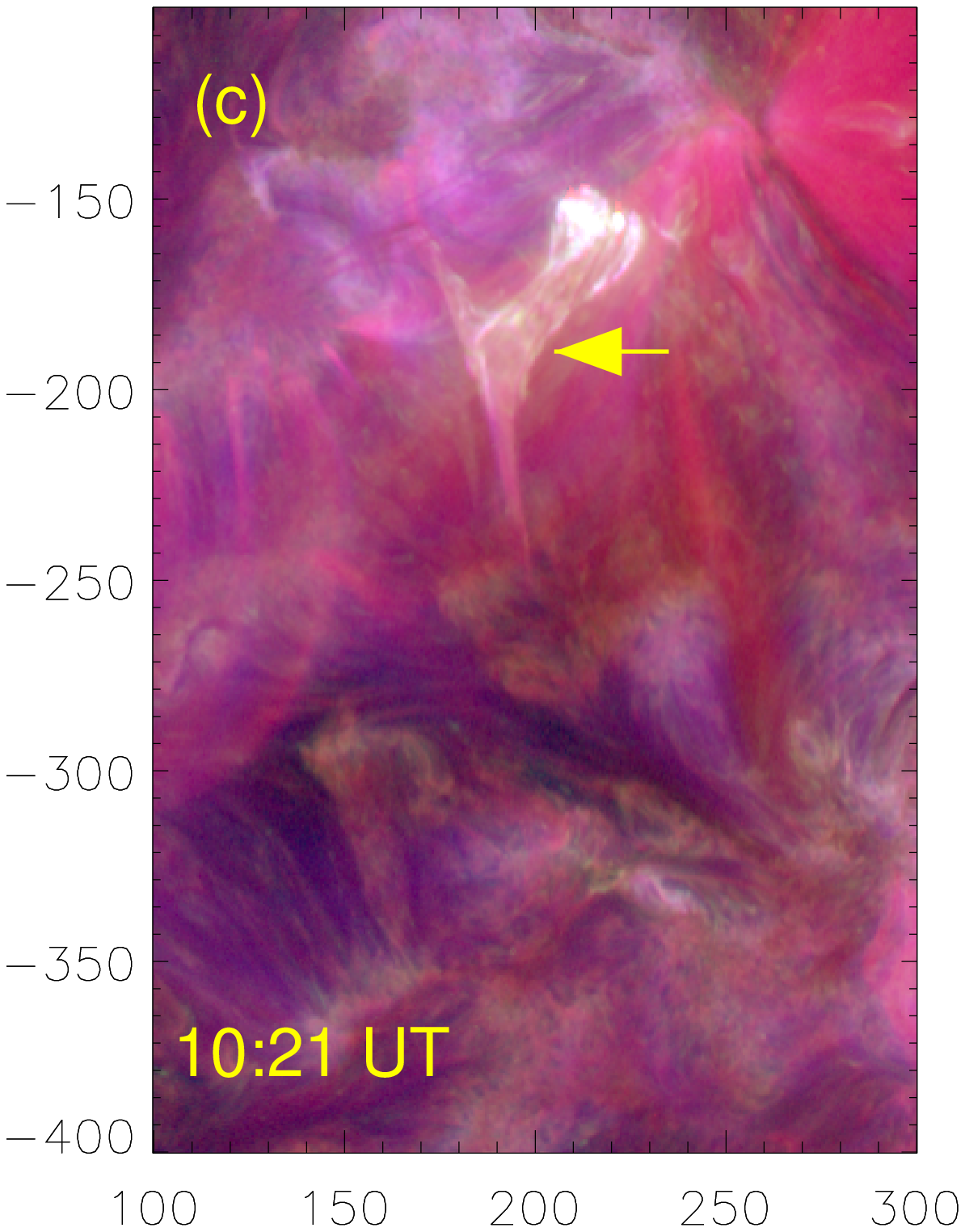}
\hspace{-3.0cm}
\includegraphics[width=6.0cm,height=5cm]{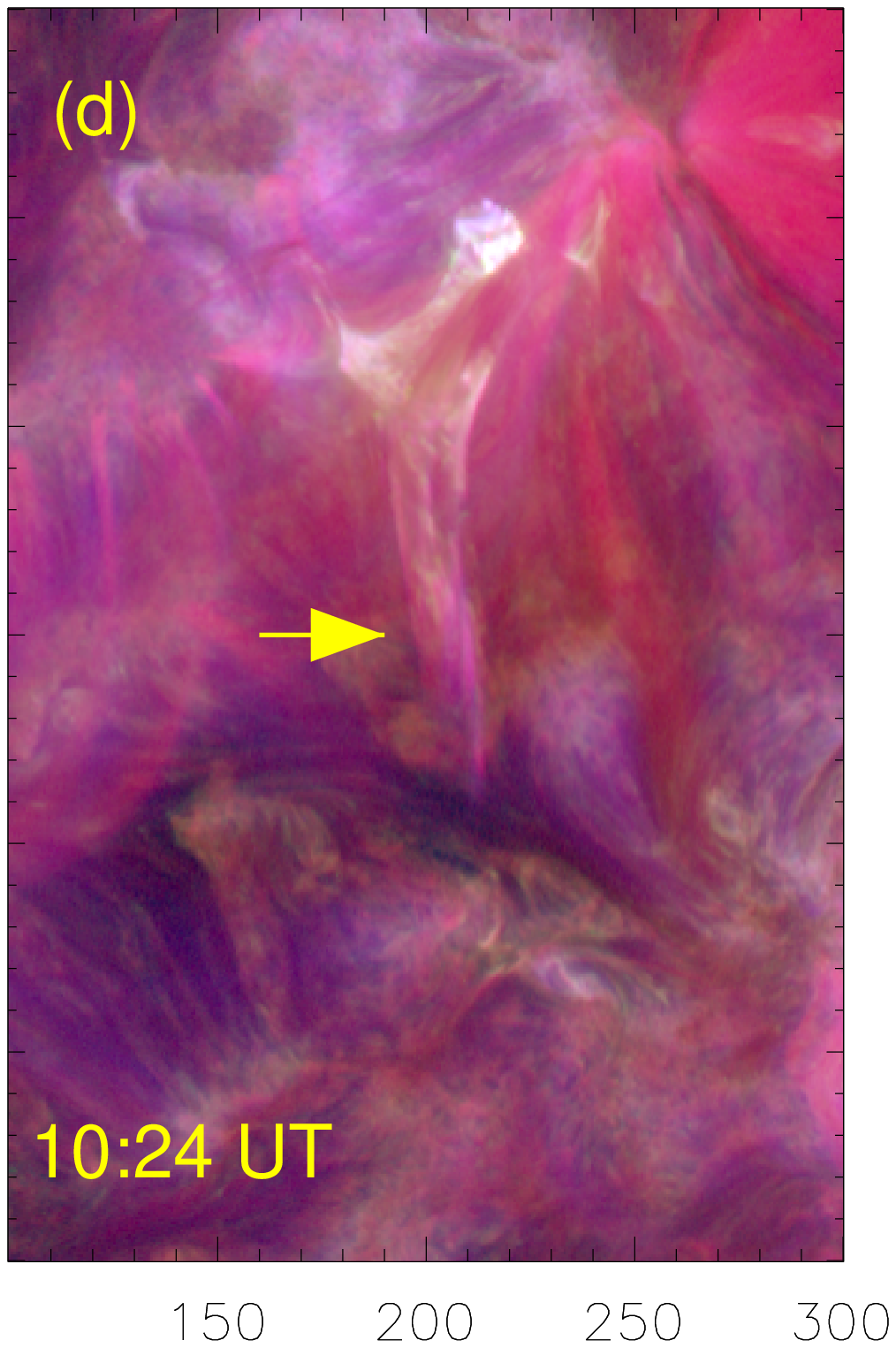}}
\vspace{-0.2cm}
\caption{SDO/AIA Composite images in 304\AA~, 171\AA~, 211\AA~ show the two-stage jet eruption associated with C-1.4 class solar flare.}
\end{center}
\end{figure*}
\section{Results and Discussion}
Fig.1 shows HINODE/XRT image at flare peak time (10:21 UT) on 2014-10-04, which demonstrates that C-1.4 class compact flare has an energy release in a very localized region on the south-east bouldary of AR 12178 outside its major sunspot. 
We have plotted GOES light curve for duration of one hour from 10:00UT to 11:00UT in soft x ray (1.0-8.0 \AA~) showing the temporal evolution of  flare Xray emissions. GOES flux peaks out at 10:21 UT indicating flare maximum. Total time duration of this flare is 8 min between 10:16 UT to 10:24 UT. Fig.2 shows a two-stage jet eruption in the composite image of SDO/AIA at 10:24 UT. The first stage of jet is erupted from the same location where this flare triggers. Basically right footpoint of this jet and C-1.4 Class flare are found to be co-spatial. The flare maximum concides with the initiation time of the first stage of jet eruption. In the second stage, the eruptive jet above the flaring region further interacts with the overlying distinct magnetic field domain in its vicinity to create the typical coronal jet moving outwards along spine field lines. The triggering mechanism of this two stage jet eruption puts rigid constraint on the existing models, and its detailed analysis is under progress by Solanki et al. (2018). 

\end{document}